\documentclass[a4paper]{ring}
\usepackage{natbib}
\usepackage{graphicx}
\usepackage[a4paper]{hyperref}
\idline{10006}{1}
\def\arcm{\hbox{$^\prime$}}
\def\arcs{\arcm\hskip -0.1em\arcm}

\begin{document}
   \title{The XMM-Newton Slew Survey
}

   \author{A.M. Read \inst{1}, \\
           R.D. Saxton\inst{2} , M.P. Esquej \inst{2}, M.J. Freyberg \inst{3}
          \and
           B. Altieri \inst{2}
}

   \offprints{A.M. Read}
\mail{Dept. of Physics and Astronomy,
              Leicester University, 
              Leicester LE1 7RH, U.K. }

   \institute{Dept. of Physics and Astronomy, Leicester University,
              Leicester LE1 7RH, U.K. \email{amr30@star.le.ac.uk}\\
              \and European Space Agency (ESA), European Space
              Astronomy Centre (ESAC), Spain \\ \and Max-Planck-Institut
              f\"{u}r extraterrestrische Physik, Giessenbachstrasse 1,
              85784 Garching, Germany }

   \authorrunning{A.M. Read et al.}  
   \titlerunning{The XMM-Newton Slew Survey} 
   \maketitle
%

\section{Introduction}

The current (April 2005) XMM archive contains $\sim$374 slew exposures
which give a uniform coverage over $\sim$10,000 square degrees
($\sim$25\% of the sky). The exposures use the EPIC medium filters,
with the observing mode set to that of the previous
observation. Average slew length is $\sim$70$^{\circ}$, and the slew
speed is about 90$^{\circ}$ per hour, i.e. the on-source time is
$\approx$14\,s (uncorrected for vignetting).

While potentially of great interest, the shallow nature of the
observations, the potential contamination by optical stars and the
contribution made by previous all-sky surveys, in particular that of
ROSAT \citet{voges}, made it unclear whether the XMM slew datasets
would provide significant new scientific results. Here we describe the
results of pilot studies, the current status of the XMM-Newton Slew
Survey and the results that have arisen.

\section{Preliminary Pilot Studies}

In obtaining calibrated event files from the EPIC slew datasets, a
small SAS OAL change was necessary. Furthermore, the normal tangential
projection used in the SAS is not valid over a whole slew, and
consequently the slews needed to be subdivided into 1 degree$^{2}$
images to maintain astrometry. Whereas sources detected in the MOS
cameras are extended into an unusable, long 4\arcm\ streak, due to the
2.6\,s frame time, the short frame time of the pn camera gives rise to
extents (essentially the slew PSF) that are not noticeable (6\arcs\
for pn FF mode, 18\arcs\ for pn eFF mode). For this reason plus the
additional MOS background, it was concluded that just the pn slew data
would be analysed.

\begin{figure}
   \centering
   \resizebox{\hsize}{!}{\rotatebox[]{0}{\includegraphics[clip=true]{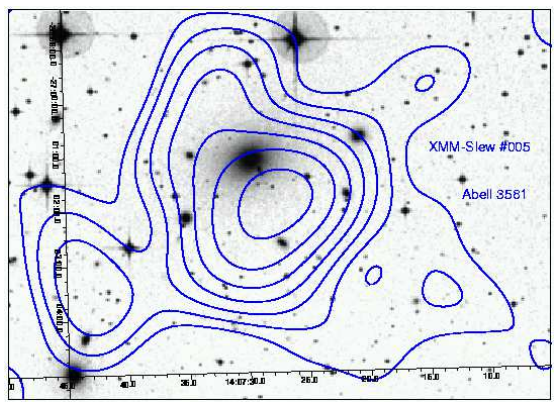}}}
     \caption{Contours of X-ray emission from a single slew across the
               galaxy cluster Abell 3581, superimposed on a DSS
               image. There is only $\approx$10\,s of on-source
               exposure here.}
        \label{A3581}
    \end{figure}

A first study of 9 slew datasets yielded $\approx$0.5 sources per
sq.\,degree to a detection likelihood threshold of 10. Of these, 10\%
appeared associated with bright stars, 45\% had ROSAT All-Sky Survey
(RASS) counterparts, and $\approx$35\% were unidentified. We also
found that we are sensitive to source extension in the brighter
sources (e.g.\,see Fig.\ref{A3581}).

A new operations strategy was put into place after this study: MOS
slews are to be used only for (closed) calibration; All pn slews
larger than 900\,s are to be down-linked and processed; For FF, eFF
and LW modes, medium filter shall be used (otherwise, closed filter).

A second pilot study investigated the optimum processing and
source-search strategies. Via SAS and other changes, we were able to
create and use correct exposure maps $-$ these produced no unusual
effects, though uneven (and heightened) slew exposure is observed at
the start/end of slews (the `closed-loop'). The optimum
source-searching strategy was found to be usage of a semi-standard
`eboxdetect (local) + esplinemap + eboxdetect (map) + emldetect'
method, tuned to $\sim$zero background, and performed on a single
image containing 0.2$-$0.5\,keV singles (pattern 0) plus
0.5$-$12.0\,keV pattern 0$-$4 events. This resulted in the largest
numbers of detected sources, whilst minimising the numbers of spurious
sources due to detector anomalies. Again, the source density was found
to be $\approx$0.5 sources per sq.\,degree.

\section{Current Status and Results}

Initial processing and event file creation has been performed for all
available 374 slews (including 206 FF, 61 eFF, 30 LW). For 54 of these
slews (31 FF, 19 eFF, 4 LW), images and exposure maps have been
created and source-searched. This has been done in 3 separate bands:
full band (0.2$-$0.5\,keV [pattern=0] + 0.5$-$12.0\,keV [pattern=0$-$4]),
soft band (0.2$-$0.5\,keV [pattern=0] + 0.5$-$2.0\,keV [pattern=0$-$4]), and
hard band (2.0$-$12.0\,keV [pattern=0$-$4]). 780 sources have been
detected in the total band, 645 in the soft band, and 96 in the hard
band. Furthermore, at the faint end, 68 sources are detected only
in the soft band, and 20 sources are detected only in the hard
band. The total of 868 sources in 1800 sq.\,degrees is again
$\approx$0.5 sources per sq.\,degree.

Many different processing problems currently affect several of the
slews. Though concentrating on slews which process cleanly, we are
finding solutions to individual problems as we progress. Some slews
contain sections with high exposure, related to the closed-loop slew
phase, and these sub-images are currently excluded. Also
high-background (flaring) slews ($\sim$25\%) are excluded at present,
though we can probably recover a lot of this data by GTI
subsetting. At present, we are source-searching 219 slews and expect
to find $\sim$3600 sources.

\begin{figure}
   \centering
   \resizebox{\hsize}{!}{\rotatebox[]{0}{\includegraphics[clip=true]{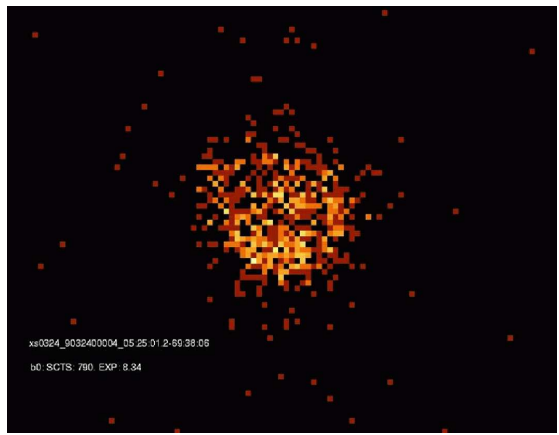}}}
     \caption{XMM slew X-ray emission from N132D.}
        \label{n132d}
    \end{figure}

A great variety of sources have already been detected, including
stars, galaxies, both interacting and normal, AGN, clusters, and SNR
(e.g. N132D $-$ see Fig.\,\ref{n132d}), plus other unusual sources,
that we know, as yet, nothing about, and extremely bright sources,
with (at present) up to 350\,ct s$^{-1}$. A large variation in source
hardness is also seen. Furthermore, several sources are detected in
more than one slew, yielding variability information. One source, so
far detected in three separate slews, appears to have varied in flux
by a factor of $\sim$2.

Again, $\sim$10\% of sources appear associated with bright stars. This
does not appear to be due to optical loading however, as no
correlation is observed between source counts and optical magnitude
for the bright stars so far detected as slew sources. This is
consistent with predictions that 5 counts are expected above 200\,eV
only for stars brighter than $V=3.75$ $-$ hence, the vast majority of
the stellar slew sources appear to to be genuine X-ray sources.

Correlations of the obtained sky positions both with 2MASS and with
RASS indicate that the pointing accuracy of the slew is very good $-$
$\sim$6\arcs, but that there is an additional second type of
positional error. This is an attitude-related error of 0$-$60\arcs\
(mean 30\arcs), but only in the slew direction, and this results in a
thin, slew-oriented `error ellipse' around each source. We believe we
may be able to remove the second slew-attitude error by re-processing
the attitude data. The $\sim$6\arcs\ error will remain of course, but
this is easily small enough to allow good optical follow-up.

\begin{figure}
   \centering
   \resizebox{\hsize}{!}{\rotatebox[]{270}{\includegraphics[clip=true]{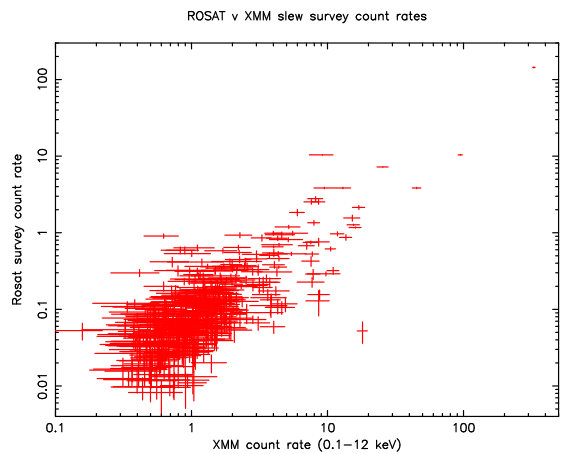}}}
     \caption{RASS (0.1$-$2.4\,keV) count rate versus XMM Slew
     (0.2$-$2.0\,keV) (pn) count rate for 407 matches. }
        \label{crate}
    \end{figure}

Approximately 60\% of the non-extended slew sources have RASS
counterparts.  Fig.\,\ref{crate} shows the RASS count rate versus the
XMM Slew count rate for the current matches. The scatter in count rate
ratio may be due to several factors, including genuine source
variability. It is seen that the harder slew sources are generally not
observed in the RASS $-$ the slew sources with RASS counterparts are
on average spectrally softer, whereas the slew sources without RASS
counterparts are on average spectrally harder.

\begin{figure}
   \centering
   \resizebox{\hsize}{!}{\rotatebox[]{0}{\includegraphics[clip=true]{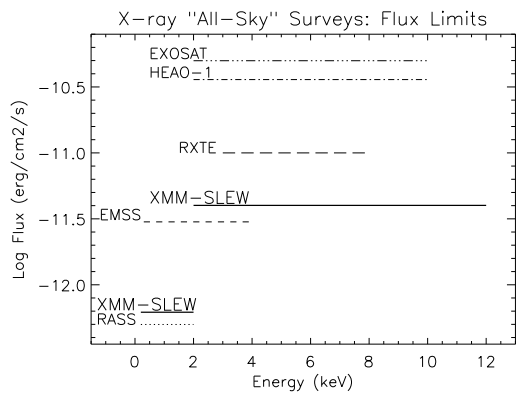}}}
     \caption{X-ray `all-sky' flux limits: Logarithm of the limiting
flux and the relevant X-ray band.  RASS: 5.0$\times10^{-13}$\,erg
cm$^{-2}$ s$^{-1}$, EMSS: 3.0$\times10^{-12}$\,erg cm$^{-2}$ s$^{-1}$
(but only 2\% of the sky), HEAO-1: 3.6$\times10^{-11}$\,erg cm$^{-2}$
s$^{-1}$, Exosat: 5.0$\times10^{-11}$\,erg cm$^{-2}$ s$^{-1}$, RXTE:
1.0$\times10^{-11}$\,erg cm$^{-2}$ s$^{-1}$ (but with only 1$^{\circ}$
positional accuracy).  }
        \label{fluxlimits}
    \end{figure}

The soft and hard band XMM-Slew survey flux limits,
6.2$\times10^{-13}$\,erg cm$^{-2}$ s$^{-1}$ (soft band) and
4.0$\times10^{-12}$\,erg cm$^{-2}$ s$^{-1}$ (hard band), are shown,
together with other `all-sky' flux limits in
Fig.\,\ref{fluxlimits}. The soft band limit is similar to the RASS
limit, and the hard band limit is very much deeper than any other
all-sky survey.

Initial processing of the slew data is now finished. The creation of
images and exposure maps, and the source-searching of these is
ongoing, as is the creation of the catalogues, the creation of DSS
images, the checking for and flagging of spurious sources and bright
stars, the calculating of hardness ratios and the cross-correlating
with RASS and other catalogues. It is envisaged that a final catalogue
will be created and ingested into the XMM-Newton XSA (at ESAC) by the
end of 2005. We are also hoping to begin investigating extended
sources and searching co-added slews in the near future.

\section{Concluding Remarks}

The XMM archive currently contains slew exposures which give a uniform
coverage of $\sim$10,000 square degrees. Analysis of a subset of these
data has revealed that thousands of entirely new sources and perhaps
new classes of sources will be discovered.

All available slew datasets have undergone initial processing. These
data cover $\sim$25\% of the sky, and at the current rate, XMM-Newton
should have completed an all-sky slew survey by 2012. To date, 54
slews (covering $\approx$4.5\% of the sky) have been source-searched
using the optimum strategies. We are detecting $\sim$0.5 sources per
sq.\,degree (to a detection likelihood of 10 [$\approx3.9\sigma$]).

The current slew survey soft band (0.2$-$2.0\,keV) detection limit is
close (to within $\sim$20\%) of the ROSAT All-Sky Survey (RASS)
limit. The hard band (2.0$-$12.0\,keV) detection limit is the deepest
ever, going more than 10 times deeper than Exosat and HEAO-1, and over
2.5 times deeper than RXTE (which only has 1$^{\circ}$ positional
accuracy anyway).

The XMM slew positional accuracy appears very good ($\approx$6\arcs),
but there exists an additional 1-D attitude error (of 30\arcs\ mean),
along the slew direction. This error may be removable via re-analysis
of the existing attitude data. We further aim to recover science from
slews affected by times of high background. We are currently on
schedule to have a final catalogue by the end of 2005.


\bibliographystyle{aa}

\end{document}